\documentstyle[tighten,floats,aps,epsf]{revtex}

\begin{document}
\title
{Structure of Excited States of $^{10}$Be studied with Antisymmetrized 
Molecular Dynamics}

\author{Y. Kanada-En'yo}

\address{Institute of Particle and Nuclear Studies, \\
High Energy Accelerator Research Organization,\\
3-2-1 Midori-cho Tanashi Tokyo 188-8501, Japan}

\author{H. Horiuchi and A. Dot\'e}

\address{Department of Physics, Kyoto University, Kyoto 606-01, 
Japan}

\maketitle
\begin{abstract}
We study structure of excited states of $^{10}$Be with the method of 
variation after
spin parity projection in the frame work of antisymmetrized molecular
dynamics.
Present calculations describe many excited states and
reproduce the experimental data of $E2$ and $E1$ transitions and the new data
of the $\beta$ transition strength successfully.
We make systematic discussions on the molecule-like structures of 
light unstable nuclei
and the important role of the valence neutrons based on the results 
obtained with the framework which is free 
from such model assumptions as the existence of inert cores and clusters.
\end{abstract}


\section{Introduction}

Owing to the radioactive nuclear beams, the data of unstable nuclei are 
increasing rapidly. Recently the structure of 
excited states as well as the ground states are very attractive in the
study of unstable nuclei.
In the light nuclear region, one of the hot and important 
subjects has been the
clustering features. The
clustering structures of the ground states
are already well known to be developed in some light ordinary nuclei such as
$^7$Li, $^8$Be and $^{20}$Ne.
In the studies of the light unstable nuclei
\cite{SEYA,ENYOb,ENYOc,TAKAMI,DOTE,ITAGAKI,OGAWA}
the clustering structures were predicted  
also in the very neutron-rich nuclei.
It is natural that various molecule-like 
states may appear
in the excited states of light unstable nuclei
because the excitation due to relative motion between 
clusters is important in the light nuclear region.
The clustering structure  
must be one of the important aspects
to understand the exotic features of
the unstable nuclei.
However there remain many mysterious problems of
clustering development and the role of valence nucleons 
in the unstable nuclei.
Our aim is to make systematic study of structure change with the
increase of the excitation energy.
We are to find new features of unstable nuclei such as exotic shapes
and molecule-like structures in the excited states and to 
understand the mechanism of the various structures.

$^{10}$Be, one of the challenges in the study of 
light unstable nuclei,
were and are studied hard theoretically
\cite{ENYOb,TAKAMI,DOTE,ITAGAKI,OGAWA,OERTZEN}.
Recent experiments of the charge exchange 
reactions $^{10}$B($^3$He,$t$)$^{10}$Be \cite{FUJIWARA}
let us know the strength of the Gamow-Teller transitions to 
the excited states of $^{10}$Be. These new data of $\beta$ transition
strength which are deduced from the cross sections at the forward angle
are very helpful to study the structure of the excited states.

In the theoretical researches on the structure of the ground states 
of Be isotopes such as the reference \cite{ENYOb}, it was found that 
the developed $2\alpha$ clustering structure in $^8$Be weakens in the 
$^{10}$Be because of the excess two neutrons. 
As for the excited states of $^{10}$Be, the previous
calculations \cite{ENYOdoc}
with a simplest version (variation after parity projection)
of antisymmetrized molecular dynamics (AMD) 
\cite{ENYOb,ENYOc,ONOa,ENYOa} 
predicted that the largely deformed states with the $2\alpha$+$2n$ 
clustering structure 
construct a rotational band $K=1^-$ in low energy region 
which is supported by the experimental energy levels.

The problem of the abnormal spin parity $1/2^+$ in the 
ground state of $^{11}$Be is considered to have a close relation with 
the deformed non-normal parity states of $^{10}$Be.
In the simple shell model consideration the normal parity of $^{11}$Be 
is negative because of the valence neutron in the $p_{1/2}$ shell, however, 
the ground states is known to be $J^\pi=1/2^+$.
It is suggested that one of the reasons for the parity inversion
 in $^{11}$Be is
the energy gain of the $sd_{1/2}$ mixing orbit in the prolately deformed 
system due to the developed clustering.
The $sd_{1/2}$ orbit of neutrons in neutron-rich Be isotopes is 
interesting and essential to understand the structure of light neutron-rich
 nuclei and is easily connected with the deformed 
structure of the non-normal parity states of $^{10}$Be.
Furthermore it is natural that the negative parity band due to the
deformed states of $^{10}$Be
makes us imagine a corresponding positive parity band 
constructed by the analogous deformed intrinsic state
where the neutrons in the $sd_{1/2}$ orbit play important roles.

Recently W. Von Oertzen proposed a dimer model \cite{OERTZEN} to describe the 
excited states of Be and B isotopes systematically. In his idea 
he supposed $2\alpha$ clusters and surrounding nucleons in
molecular orbits like $\sigma$ and $\pi$ bonds as the orbits of electrons
in the molecule. 
The neutron-rich Be nuclei can be written by $2\alpha$ and neutrons which 
occupy the molecule-like $\sigma$ and $\pi$ orbits.
This idea is helpful to understand the level structure of 
many excited states of $^9$Be and $^{10}$Be and to study 
the mechanism of the deformation of Be isotopes.

A. Dot\'e et al. studied Be isotopes
by microscopic calculations with 
the simplest version of AMD under a constraint on the deformation parameter 
\cite{DOTE}.
They analyzed the single-particle orbits of the valence neutrons 
and found the importance of molecular orbits.
It was found that in the negative parity state of $^{10}$Be 
the main component of the
valence neutrons is the positive parity orbit
which seems to be the $\sigma$ orbit in terms of the dimer model.
They tried to represent the excited $0^+_2$ state  
of $^{10}$Be by constructing 2p-2h states using the wave function of the 
negative parity state 
which is considered to be 1p-1h state. 
Comparing the optimum deformation parameters for three states
$0^+_1$, $1^-$ and $0^+_2$, 
they found that the intrinsic deformation is larger in $1^-$ 
than $0^+_1$ and largest in $0^+_2$. The results 
is consistent with the ones predicted by W. Von Oertzen.
They calculated excited states $0^+_2$ of $^{10}$Be 
without assuming any cluster cores 
for the first time and confirmed that the 
$2\alpha$ cluster cores may exist in the excited states.
However since the total-angular-momentum projection was 
not done microscopically, 
the discussion didn't go beyond the intrinsic system and 
it is not easy to study other excited states.
 They have not mentioned the data such as $E1,E2$ and
$\beta$ transitions which are important in the study of excited states.

$^{10}$Be has been studied with other microscopic models such as multi
cluster models \cite{OGAWA}, AMD+GCM(Generator Coordinate Method)
\cite{ITAGAKI} and 
the parity-projected Hartree-Fock(HF) calculations \cite{TAKAMI}.
In the former two method the existence of $\alpha$ clusters was assumed.
In the last work with parity-projected HF calculations they assumed the
excited $0^+_2$ state as a 2p-2h state by choosing the 
positive parity eigen state for the valence two neutrons. 
In the work with HF calculations,
the analysis was limited in the intrinsic system and it is not easy to 
study the higher excited states.

Our aim is to make systematic studies on the structure of many
excited states of the light unstable nuclei with a microscopic model which is 
free from such assumptions 
as the inert cores, the existence of clusters, and 
the particle-hole configurations.
For this aim we adopt a method of 
antisymmetrized molecular dynamics(AMD).
AMD has been already proved to be a very useful theoretical 
approach for the structure of the light nuclei 
\cite{ENYOb,ENYOc,DOTE,ENYOa,ENYOe}.
In the AMD framework basis wave functions of the system are written by 
Slater determinants where the spatial part of each 
single-particle wave function is a Gaussian wave packet.
In the previous works on the structure of 
Li, Be and B isotopes \cite{ENYOb,ENYOc},
we applied the simplest version of AMD in which the energy variation is
done after the parity projection but before the
total-angular momentum projection. 
It was found that the structure of the ground state changes rapidly between
the shell-model-like structure and the clustering structure
as the neutron number increases up to the neutron-drip line.
The AMD calculations succeeded to reproduce well
the experimental data for electro-magnetic properties such as the 
magnetic dipole moments and the electric quadrupole moments. 
It owes to the flexibility of 
the AMD wave function which can represent the clustering structures
of light unstable nuclei without assuming any inert cores and clusters.

In the present paper,
we study the structure of the excited states of $^{10}$Be
by performing variational calculations after the spin parity projection
with finite-range interactions in the framework of AMD.
For higher states, AMD wave functions are superposed so as to be orthogonal
to the lower states. The AMD approach of the variation after spin parity 
projection(VAP) has been already found to be
advantageous for the study of excited states of 
the light nuclei and be very useful to describe various 
molecule-like structures \cite{ENYOe,ENYOf} .
We do not rely on any assumptions such as the inert cores, 
the single-particle orbits in the mean field, and the existence of 
clusters. 
By microscopic calculations of the expectation values of the corresponding
operators,
we can acquire the theoretical values of
$E2,E1$ and  $\beta$ transitions which are the valuable informations 
of the excited states to be compared directly with the experimental data. 

In Sec.\ \ref{sec:formulation}, we explain 
the formulation of AMD for the study of 
the nuclear structure of excited states.
The effective interactions are described in Sec.\ \ref{sec:inter}, and
the results are presented in Sec.\ \ref{sec:results} comparing with the
experimental data. In Sec.\ \ref{sec:discussion}, 
the detailed discussions on the 
structures and on the mechanism of the development of 
the molecule-like structures are made. Finally summary is given in 
Sec.\ \ref{sec:summary}.

\section{Formulation}
 \label{sec:formulation}

In this section we explain the formulation of AMD 
for the study of nuclear structure of the excited states 
\cite{ENYOe,ENYOf}.

\subsection{Wave function}
The wave function of a system is written by AMD wavefunctions,
\begin{equation}
\Phi=c \Phi_{AMD} +c' \Phi '_{AMD} + \cdots .
\end{equation}
An AMD wavefunction of a nucleus with mass number $A$
is a Slater determinant of Gaussian wave packets;
\begin{eqnarray}
&\Phi_{AMD}({\bf Z})=\frac{1}{\sqrt{A!}}
{\cal A}\{\varphi_1,\varphi_2,\cdots,\varphi_A\},\\
&\varphi_i=\phi_{{\bf X}_i}\chi_{\xi_i}\tau_i :\left\lbrace
\begin{array}{l}
\phi_{{\bf X}_i}({\bf r}_j) \propto
\exp\left 
[-\nu\biggl({\bf r}_j-\frac{{\bf X}_i}{\sqrt{\nu}}\biggr)^2\right],\\
\chi_{\xi_i}=
\left(\begin{array}{l}
{1\over 2}+\xi_{i}\\
{1\over 2}-\xi_{i}
\end{array}\right),
\end{array}\right. 
\end{eqnarray}
where the $i$-th single-particle wave function $\varphi_i$
is a product of the spatial wave function, the intrinsic spin function and 
the isospin function. The spatial part $\phi_{{\rm X}_i}$ is presented by 
complex parameters $X_{1i}$, $X_{2i}$, $X_{3i}$,
$\chi_{\xi_i}$ is the intrinsic spin function parameterized by
$\xi_{i}$, and $\tau_i$ is the isospin
function which is fixed to be up(proton) or down(neutron)
 in the present calculations.
Thus an AMD wave function is parameterized by a set of complex parameters
${\bf Z}\equiv \{X_{ni},\xi_i\}\ (n=1,3\ \hbox{and }  i=1,A)$, 
where ${\bf X}_{i}$'s indicate the centers of 
Gaussians of the spatial part
and $\xi_{i}$'s are the parameters for 
the directions of the intrinsic spins.

If we consider a parity eigen state projected from a AMD wave function
the total wave function consists of two Slater determinants,
\begin{equation}
\Phi({\bf Z})=(1\pm P) \Phi_{AMD}({\bf Z}),
\end{equation}
where $P$ is a parity projection operator.
In case of a total-angular momentum eigen state
 the wave function of a system
is represented by integral of the rotated states,
\begin{equation}
\Phi({\bf Z})=P^J_{MK'}\Phi_{AMD}({\bf Z}) = 
\int d\Omega D^{J*}_{MK'}(\Omega)R(\Omega)\Phi_{AMD}({\bf Z}),
\end{equation}
for which the expectation values of operators are numerically  calculated by 
a summation of mesh points on the Euler angles $\Omega$. 

In principal the total wave function can be superposition of independent
AMD wave functions. In order to construct higher excited states
we consider superposition 
of the spin parity projected AMD wave functions $P^{J\pm}_{MK'}\Phi_{AMD}$, 
\begin{equation}
\Phi=cP^{J\pm}_{MK'}\Phi_{AMD}({\bf Z})
+c'P^{J\pm}_{MK'}\Phi_{AMD}({\bf Z}')+\cdots.
\end{equation}
The detail is mentioned later in subsection \ref{subsec:excited}.

\subsection{Energy variation}
We make variational calculations
to find the state which minimizes the energy of the system;
\begin{equation}
\frac{\langle\Phi|H|\Phi\rangle}{\langle\Phi|\Phi\rangle}
\end{equation}
by the method of frictional cooling.
Regarding the frictional cooling method in AMD, the reader is referred to 
papers \cite{ENYOb,ENYOa}.
For the wave function $\Phi({\bf Z})$ 
parameterized by ${\bf Z}$, 
the time development of the parameters are given by the frictional cooling
equations,
\begin{equation}
\frac{dX_{n k}}{dt}=
(\lambda+i\mu)\frac{1}{i \hbar} \frac{\partial}{\partial X^*_{n k} }
\frac{\langle \Phi({\bf Z})|H|\Phi({\bf Z})\rangle}{\langle \Phi({\bf Z})
|\Phi({\bf Z})\rangle},
\quad (n=1,3\quad k=1,A)
\end{equation}
\begin{equation}
\frac{d\xi_{k}}{dt}=(\lambda+i\mu)\frac{1}{i\hbar}
\frac{\partial}{\partial\xi^*_{k}}
\frac{\langle \Phi({\bf Z})|H|\Phi({\bf Z})\rangle}{\langle \Phi({\bf Z})
|\Phi({\bf Z})\rangle},
\quad (k=1,A)
\end{equation}
with arbitrary real numbers $\lambda$ and $\mu < 0$. It is easily proved that 
 the energy of the system decreases with time. After sufficient time
steps of cooling, 
the parameters for the wave function of 
minimum-energy state are obtained. 

\subsection{Lowest $J^\pm$ states}
In order to obtain the wave function for the lowest $J^\pm$ state,
we perform the energy variation for the spin parity eigenstates projected
from an AMD wave function. In this case the trial function is 
$\Phi=P^{J\pm}_{MK'}\Phi_{AMD}({\bf Z})$. 
In the previous works \cite{ENYOb,ENYOc} with the simplest version of AMD 
for the study of nuclear structure, the approach was variation after
only the parity projection but variation before 
the total spin projection(VBP).
In the present paper,  
the approach is variation after the spin parity projection(VAP).
First we make VBP calculations to prepare
an initial wave function
 $\Phi_{AMD}({\bf Z}_{init})$ for the VAP calculations.
We choose an appropriate $K'$ quantum that gives the minimum 
diagonal energy of the spin parity eigenstate
$\langle P^{J\pm}_{MK'}\Phi_{AMD}({\bf Z}_{init})|H|P^{J\pm}_{MK'}
\Phi_{AMD}({\bf Z}_{init}) \rangle/ 
\langle P^{J\pm}_{MK'}({\bf Z}_{init})|P^{J\pm}_{MK'}({\bf Z}_{init})
 \rangle$,
where $K'$ is the component of the total-angular momentum along
the approximately principal axis in the intrinsic system.
For each spin parity $J^\pm$, we perform VAP calculation 
for $\langle P^{J\pm}_{MK'}\Phi_{AMD}({\bf Z})|H|P^{J\pm}_{MK'}
\Phi_{AMD}({\bf Z}) \rangle/ 
\langle P^{J\pm}_{MK'}({\bf Z})|P^{J\pm}_{MK'}({\bf Z})
 \rangle$ with the adopted $K'$
quantum from the initial state. In the VAP procedure, the principal $z$-axis 
of the intrinsic deformation is not assumed to equal with the
 $3$-axis of Euler angle in the total-angular momentum
projection. In general the principal $z$-axis is automatically 
determined in the energy variation, that is to say that the
obtained state $P^{J\pm}_{MK'}\Phi_{AMD}$ by VAP with 
a given $K'=\langle J_3\rangle$ can be the state with 
so-called $K$-mixing
in terms of the intrinsic deformation where the 
$K=\langle J_z\rangle$ quantum is defined on the principal axis.
In many cases, $z$-axis determined in the VAP
calculation is found to be approximate to the $3$-axis.
It means that
the obtained states do not contain $K$-mixing so much, and 
$K'$ is considered to be the approximate $K$
quantum for the principal axis.
The excited $J^\pm$ states
in the band $K^\pi=K''^\pm$ other than $K'$ quantum  
of the lower $J^\pm$ states are obtained by VAP calculations for 
$P^{J\pm}_{MK''}\Phi_{AMD}$ under the constraint on 
the principal axis $z$
as equal to the $3$-axis.  

\subsection{Higher excited states \label{subsec:excited}}
As mentioned above,
with the VAP calculation for $\Phi({\bf Z})=P^{J\pm}_{MK'}\Phi_{AMD}({\bf Z})$
of the $J^\pm$ eigen state with $K'$,
we obtain the set of parameters ${\bf Z}={\bf Z}^{J\pm}_1$ which present
the wave function for the first $J^\pm$ state. 
To search the parameters ${\bf Z}$  for the higher excited $J^\pm$ states
in the $n$-th $K^\pi=K'^\pm$ band, 
the wave functions are superposed to be orthogonal
to the lower states as follows. The parameters ${\bf Z}^{J\pm}_n$ 
for the $n$-th
$J^\pm$ state are reached by varying the energy of the orthogonal
component to the lower states;
\begin{equation}
\Phi({\bf Z})=P^{J\pm}_{MK'}\Phi_{AMD}({\bf Z})-\sum^{n-1}_{k=1}
{\langle P^{J\pm}_{MK'}\Phi_{AMD}({\bf Z}^{J\pm}_k)
|P^{J\pm}_{MK'}\Phi_{AMD}({\bf Z})\rangle 
\over
\langle P^{J\pm}_{MK'}\Phi_{AMD}({\bf Z}^{J\pm}_k)
|P^{J\pm}_{MK'}\Phi_{AMD}({\bf Z}^{J\pm}_k)\rangle} 
P^{J\pm}_{MK'}\Phi_{AMD}({\bf Z}^{J\pm}_k).\label{eqn:excite}
\end{equation}

\subsection{Expectation values}  

After VAP calculations for various $J^\pm_n$ states, the 
intrinsic states
$\Phi^1_{AMD}, 
\Phi^2_{AMD},\cdots, 
\Phi^m_{AMD}$, 
which approximately correspond to the $J^\pm_n$ states, 
are obtained as much as the number 
of the calculated levels.
Finally we determine the wave functions for the $J^\pm_n$ states
by diagonalizing the Hamiltonian matrix 
$\langle P^{J\pm}_{MK'} \Phi^i_{AMD}
|H|P^{J\pm}_{MK''} \Phi^j_{AMD}\rangle$
and the norm matrix
$\langle P^{J\pm}_{MK'} \Phi^i_{AMD}
|P^{J\pm}_{MK''} \Phi^j_{AMD}\rangle$
simultaneously with regard to ($i,j$) for 
all the intrinsic states and ($K', K''$).
In comparison with the experimental data such as energy levels and
 $E2$ transitions, the theoretical values are calculated with the 
final states after diagonalization.

\section{Interactions} 
\label{sec:inter}

The adopted interaction for the central force is the case 3 of 
MV1 force \cite{TOHSAKI},
which contains a zero-range three-body force $V^{(3)}$ 
as a density dependent 
term in addition to the two-body interaction $V^{(2)}$ of 
the modified Volkov No.1 force, 
\begin{eqnarray}
& V_{DD}=V^{(2)}+V^{(3)}\\
& V^{(2)}=(1-m+b P_\sigma-h P_\tau -m P_\sigma P_\tau )
\left\lbrace 
V_A \exp\left[-\left(\frac{r}{r_A}\right)^2\right]+ 
V_R \exp\left[-\left(\frac{r}{r_R}\right)^2\right]\right\rbrace,\\
& V_A = -83.34 \hbox{MeV}, r_A = 1.60 \hbox{fm},
 V_R = 99.86 \hbox{MeV}, r_R = 0.82 \hbox{fm},\\
& V^{(3)} = v^{(3)}\delta({\bf r}_1-{\bf r}_2)\delta({\bf r}_1-{\bf r}_3),
\ v^{(3)} = 5000 \hbox{MeV fm$^6$},
\end{eqnarray}
where $P_\sigma$ and $P_\tau$ stand for the spin and isospin exchange 
operators, respectively. 
As for the two-body spin-orbit force $V_{LS}$, we adopted the G3RS force 
\cite{LS} as follows,
\begin{eqnarray}
& V_{LS}= \left\{ u_I \exp\left(-\kappa_I r^2\right) +
u_{II} \exp\left(-\kappa_{II} r^2\right)\right\} 
\frac{(1+P_\sigma)}{2}
\frac{(1+P_\tau)}{2}
{\bf L}\cdot ({\bf S_1}+{\bf S_2}),\\
& \kappa_I = 5.0 \hbox{fm}^{-2}, 
 \kappa_{II} = 2.778 \hbox{fm}^{-2}. 
\end{eqnarray}
The Coulomb interaction $V_{{\rm C}}$ 
is approximated by a sum of seven Gaussians.

We treat  the resonance states 
within a bound state approximation by 
situating an artificial barrier ${\cal E}_{{\rm barr}}$.
In the variational calculation we variate the modified energy 
${\cal E}'={\cal E}+{\cal E}_{{\rm barr}}$ with the additional 
artificial energy ${\cal E}_{{\rm barr}}$ 
instead of ${\cal E}\equiv {\langle \Phi |H|\Phi\rangle}/
{\langle\Phi | \Phi\rangle}
$, where $H$ is the Hamiltonian operator; 
$H=T+V_{DD}+V_{LS}+V_{{\rm C}}$.
For the barrier in the present calculations
 we adopt a function of the real part of the 
centers ${\bf X}_i$ of 
the single-particle wave functions as follows,
\begin{equation}
 {\cal E}_{{\rm barr}}=V_W \sum^A_{i=1}
{\rm exp}\left[ -2\nu\left(r_a-
\frac{|{\rm Re} [{\bf X}_i]|}{\sqrt{\nu}}\right)^2\right]
\end{equation}
\begin{eqnarray}
& r_a=r_0\left (2{\rm Max}[Z,N]\right )^{1/3}+b\\
& V_W=2.5 {\ \rm MeV},\ \  r_0=1.2 \ {\rm fm},\ \  b=\sqrt{\frac{3}{2\nu}}
\  {\rm fm}, 
\end{eqnarray}
where $Z$ and $N$ are proton and neutron numbers, respectively.
We put the artificial barrier only in the variational calculation, but
do not put the barrier in the diagonalization of the Hamiltonian matrix
after VAP calculations.
In the present calculations of $^{10}$Be, 
the artificial energies due to the barrier
are found to be a few hundreds keV at most and make no significant effect
on the results in most levels.
  
\section{Results}\label{sec:results}
The structure of the excited states of $^{10}$Be is studied 
with the VAP calculations in the framework of AMD.
In this section we display the theoretical results of the excitation 
energies, $E2$, $E1$, and $\beta$ transitions
which can be directly compared with the experimental data.
The detail of the structures are discussed in the next section. 
The adopted parameters of the interactions are 
$m=0.62$, $b=h=0$ for the Majorana,
Bertlett and Heisenberg 
terms of the central force and the strength of the spin-orbit force
$u_I=-u_{II}=3000$ MeV (case(1)).
Trying another set of parameters case(2) with $m=0.65$, $b=h=0$ and 
$u_I=-u_{II}=3700$ MeV, we did not find significant differences
 in the results. The set of parameters of case(1) is the one
 adopted in the work on $^{12}$C \cite{ENYOe}. On the other hands, 
the VAP calculations with the set of interactions 
case(2) reproduce the parity inversion of the ground state 
of $^{11}$Be.
The optimum width parameters $\nu$ of wave packets are chosen to be 
0.17 fm$^{-2}$ for case(1) and 0.19 fm $^{-2}$ for case(2) 
which give the minimum energies in VBP calculations of $^{10}$Be.
The resonance states are treated within a bound state approximation 
by situating an artificial barrier out the surface as mentioned in section
\ref{sec:inter}.

The lowest $J^\pm$ states are obtained by 
VAP calculations for $P^{J\pm}_{MK'}\Phi_{AMD}$ with $(J^\pm,K')$=
$(0^+,0)$, $(2^+,0)$, $(3^+,+2)$, $(4^+,0)$, $(1^-,-1)$, $(2^-,-1)$,  
$(3^-,-1)$, $(4^-,-1)$. 
Considering $0^+_2$ state to be $0^+$ state in the second 
$K^\pi=0^+$ band, 
the $0^+_2$ state is calculated by VAP as the higher excited state orthogonal
to the lowest $0^+_1$ state as explained 
in the subsection \ref{subsec:excited}. That is to say that 
the $0^+_2$ state is obtained by VAP for $\Phi({\bf Z})$ 
in Eq. \ref{eqn:excite} with 
$(J^\pm,K',n)=(0^+,0,2)$.
In the case of higher $2^+$ states, 
we put the constraint that 
the approximately principal $z$-axis of the intrinsic 
deformation equals to the 3-axis of the Euler angle in the 
total-spin projection.
According to VBP calculations 
the second $2^+$ state is described 
as the band head of the lowest $K^\pi=2^+$ band.
Therefore we construct the state $2^+_2$ 
by choosing $(J^\pm,K')$ of  $P^{J\pm}_{MK'}\Phi_{AMD}$
as $(J^\pm,K')=(2^+,+2)$ under the constraint on the principal $z$-axis
which keeps the approximately orthogonality to the 
lowest $2^+$ state with $(J^\pm,K')=(2^+,0)$.
The third $2^+_3$ state is easily conjectured to 
be $2^+$ state in the second $K^\pi=0^+_2$ band like $0^+_2$ state.
We obtain the $2^+_3$ state by VAP for $\Phi({\bf Z})$ in Eq.\ref{eqn:excite}
with $(J^\pm,K',n)=(2^+,0,2)$ in the same way as the $0^+_2$ state, under
the constraint on the principal $z$-axis mentioned above.
It means that the orthogonal condition to $2^+_1$ 
is kept by superposing two 
wave functions 
as described in the subsection \ref{subsec:excited},  
while the orthogonality to $2^+_2$ ($K^\pi=2^+$) is taken into account
by choosing the different $K$ quantum $K'=0^+$.

\begin{figure}
\noindent
\epsfxsize=1.0\textwidth
\centerline{\epsffile{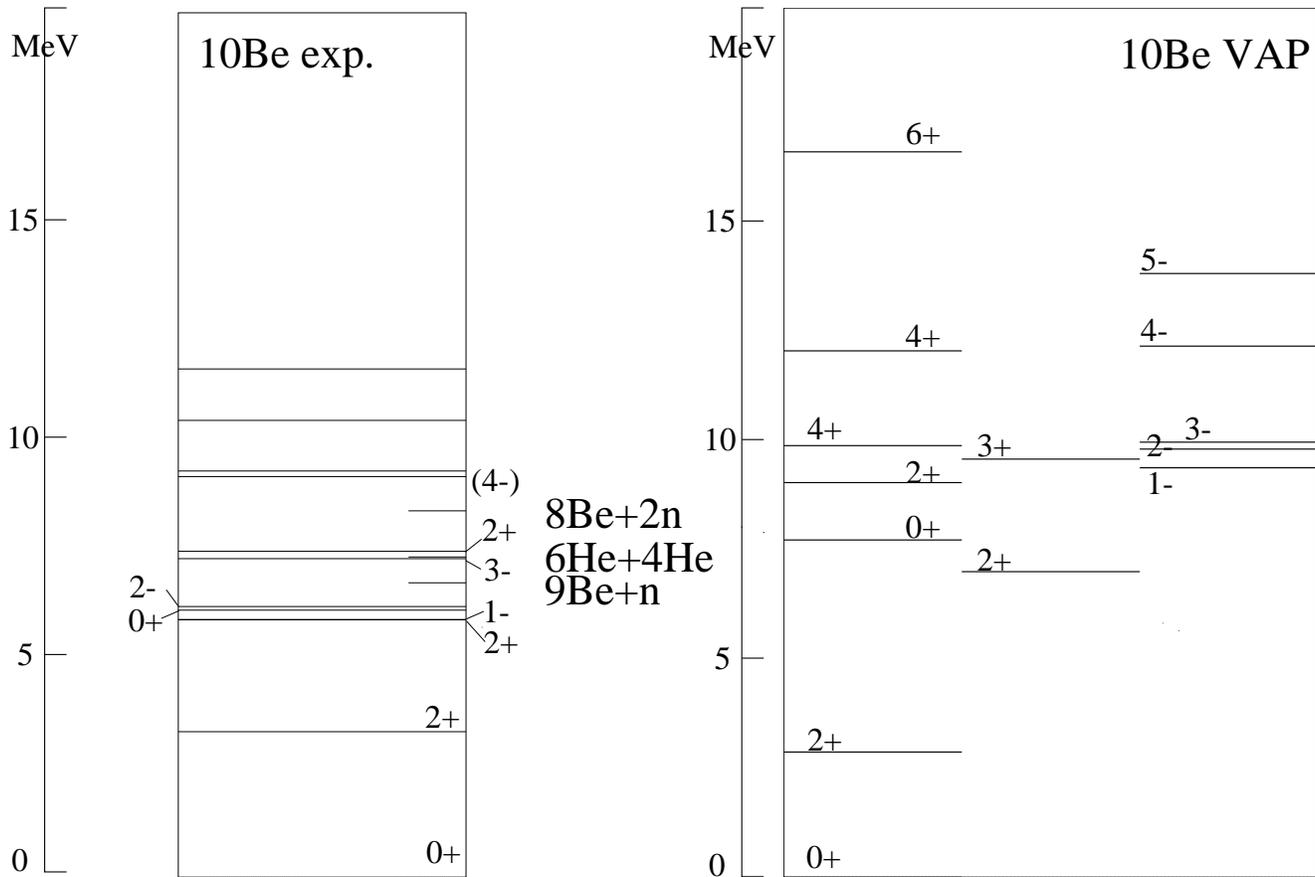}}
\caption{\label{fig:be10sped}
Excitation energies of the levels in $^{10}$Be.
Theoretical results are calculated by the diagonalization of the 
states obtained with VAP by using the interactions case(1).
}
\end{figure}

The binding energy obtained with the case(1) interactions is 61.1 MeV, 
and the one with case(2) is 61.3 MeV.
The excitation energies of the results are displayed in 
Fig.\ref{fig:be10sped}. By
diagonalization of the Hamiltonian matrix 
the excited states $4^+_2$, $6^+$ are found in the rotational band 
$K^\pi=0^+_2$ and $5^-$ 
is seen in the $K^\pi=1^-$ band.
Comparing with the experimental data, the level structure is well reproduced 
by theory. Although it is difficult to estimate the width of resonance 
within the present framework, the theoretical results suggest
the existence of $3^+$, $4^+$, $6^+$ and $5^-$ states 
which are not experimentally identified yet.
The excited levels can be classified as the rotational bands
$K^\pi=0^+_1$, $2^+$, $0^+_2$ and $1^-$. The intrinsic structures of these
bands are discussed in detail in the next section.

The data of transition strength are of great help to investigate the 
structures of the excited states. The theoretical results 
with the interaction case (1) and the experimental data
of $E2$ and $E1$ transition strength are presented in Table \ref{tab:be10e2}.
The theoretical values well agree with the experimental data.

\begin{table}
\caption{\label{tab:be10e2} $E2$ and $E1$ transition strength. The 
theoretical results with the interactions case(1) are compared with
the experimental data \protect\cite{AJZEN}.}
\begin{center}
\begin{tabular}{cccc}
 transitions & Mult. & exp.  & theory (e fm$^2$) \\
\hline
$2^+_1\rightarrow 0^+_1$ & $E2$ & 10.5$\pm$1.1 (e fm$^2$)  & 11 (e fm$^2$) \\
$0^+_2\rightarrow 2^+_1$ & $E2$ & 3.3$\pm$2.0 (e fm$^2$) & 0.6 (e fm$^2$) \\
$0^+_2\rightarrow 1^-_1$ & $E1$ & 1.3$\pm$0.6$\times 10^{-2}$ (e fm)  &
 0.6$ \times 10^{-2}$ (e fm)\\
\end{tabular}
\end{center}
\end{table}

The strength of the $\beta$ decays of Gamow-Teller(GT) type 
transitions can be deduced from the
cross sections at the $0^\circ$ forward angle 
of the charge exchange reactions which have been measured recently
\cite{FUJIWARA}. 
These new data for the Gamow-Teller type $\beta$ transitions are 
very useful probes to discuss the
structures of the excited states of unstable nuclei.
Table \ref{tab:be10beta} shows the values of $B(GT)$.
The experimental values for the $\beta$ transitions from $^{10}$B(3+) to 
$^{10}$Be$^*$ are deduced from the data of the reaction 
$^{10}$B(t,$^3$He)$^{10}$Be.
As for the theoretical values, the wave functions for the neighbor nucleus 
$^{10}$B are calculated with VAP where 
$(J^\pm, K')$ are chosen to be ($3^+,-3$) for the ground $3^+_1$ state 
and ($1^+,-1$) for the $1^+_1$ state.
$^{10}$Be and $^{10}$B are calculated with the case (1) and
(2) interactions. 
The theoretical values reasonably match to the experimental data.
The strength is not so sensitive to the interactions except for the
decay $^{10}$B$(3^+)\rightarrow ^{10}$Be($2^+_1$). 
The results of the GT transition 
$^{10}$B(3$^+$)$\rightarrow$$^{10}$Be($2^+_1$) 
with case(1) and case(2) interactions underestimate the experimental 
data, while with another set of interactions; 
Volokov No.2($m=0.38,b=0,2,h=-0.4$)+G3RS($u_I=-u_{II}=1600$MeV)+Cloumb
$B(GT)$ is calculated to be 0.41 ($g_V^2 \over 4\pi$)
which is larger than the experimental data.  
Since the data for 
$^{10}$B$(3^+)\rightarrow^{10}$Be(9.4MeV) well correspond
to the theoretical value of 
$^{10}$B$(3^+)\rightarrow^{10}$Be($3^+_1$),
it is natural to consider the excited level of $^{10}$Be at 9.4MeV
 as the $3^+_1$ state.
The result of $B(GT)$ for 
$^{10}$Be$(0^+_1)\rightarrow^{10}$B($1^+$) is consistent with
the experimental data of the $\beta$ decay from the mirror nucleus 
$^{10}$C$(0^+_1)\rightarrow^{10}$B($1^+$). 
It is suggested that the $B(GT)$ for
$^{10}$B$(3^+)\rightarrow^{10}$Be($2^+_3$) is not large 
because of the well developed clustering structure of $^{10}$Be($2^+_3$). 
We will discuss again the $\beta$ decays combining with the structures of 
excited states in the next section.
In the reference \cite{FUJIWARA} of the measurements of
the strength B(GT), they mentioned about 
the isospin symmetry violation of $\beta$ decay strength comparing 
B(GT$_-$) for $^{10}$B($3^+)\rightarrow  ^{10}$Be($2^+_2$) with
B(GT$_+$) for the mirror reaction to $^{10}$C.
In the present calculations, we could not find 
the difference between B(GT$_-$) and B(GT$_+$) because
there is little violation of  isospin symmetry
between the structures of $^{10}$Be($2^+_2$) and the mirror state
of $^{10}$C.
Taking into account of the experimental imformation that 
there exists a significant difference 
between the excitation energies of the 
daughter states of $^{10}$Be and $^{10}$C, 
the isospin violation of the structures of these
mirror states may be possible.
We need more detailed investigations to 
discuss the problem of the isospin symmetry violation of B(GT).

\begin{table}
\caption{ \label{tab:be10beta} $B(GT)$ of $\beta$ decays with 
the unit $g_V^2 \over 4\pi$.
The experimental data are the values$^{(a)}$ 
deduced from the cross sections of 
$^{10}$B(t,$^3$He)$^{10}$Be$^*$ at $0^\circ$ forward 
angle \protect\cite{FUJIWARA}
 and the 
one$^{(b)}$ from the log$ft$ value \protect\cite{NTAB}. 
The theoretical values are obtained with case (1) and case (2) interactions
for $^{10}$Be and $^{10}$B.}

\begin{center}
\begin{tabular}{ccccc}
&   & exp. &&\\ 
&initial & final & & \\
&$(J^\pi,E_x)$ (MeV) & $(J^\pi,E_x)$ (MeV) & B(GT) ($g_V^2 \over 4\pi$) & \\
\hline
&$^{10}$B($3^+$,0) & $^{10}$Be($2^+_1$,3.37) & 0.13$\pm$ 0.05$^{a)}$ &\\
&$^{10}$B($3^+$,0) & $^{10}$Be($2^+_2$,5.96) & 1.51$\pm$ 0.21$^{a)}$ &\\
&$^{10}$B($3^+$,0) & $^{10}$Be($2^+$ or $3^+$, 9.4) & 0.49$\pm$ 0.13$^{a)}$ &\\
&$^{10}$C($0^+$,0) & $^{10}$B($1^+$, 0.72) & 6.25$^{b)}$ &\\
\hline
\hline
& & theory case(1)& &\\
\hline
&$^{10}$B($3^+$) & $^{10}$Be($2^+_1$) & 0.02 &\\
& & $^{10}$Be($2^+_2$) & 1.7 &\\
& & $^{10}$Be($3^+_1$) & 0.61 &\\
& & $^{10}$Be($4^+_1$) & 0.12 &\\
& & $^{10}$Be($2^+_3$) & 0.05 &\\
&$^{10}$Be($0^+_1$) & $^{10}$B($1^+$) & 4.4 &\\
\hline
\hline
& & theory case(2)& &\\
\hline
&$^{10}$B($3^+$) & $^{10}$Be($2^+_1$) & 0.00 &\\
& & $^{10}$Be($2^+_2$) & 1.4 &\\
& & $^{10}$Be($3^+_1$) & 0.58 &\\
& & $^{10}$Be($4^+_1$) & 0.14 &\\
& & $^{10}$Be($2^+_3$) & 0.01 &\\
&$^{10}$Be($0^+_1$) & $^{10}$B($1^+$) & 3.7 &\\
\end{tabular}
\end{center}
\end{table}

\section{Discussion}
\label{sec:discussion}
In this section we discuss the structure of the excited states
by analyzing the wave functions.
Even though the states obtained by VAP mix after the diagonalization 
of the Hamiltonian matrix, the state 
$P^{J\pm}_{MK'}\Phi_{AMD}({\bf Z}^{J\pm}_n)$ projected from a Slater 
determinant obtained in VAP with $(J^\pm,K',n)$ is dominant 
in the final result of the $J^\pm_n$ state.
In this section we consider  
the Slater determinant $\Phi_{AMD}({\bf Z}^{J\pm}_{n})$ 
as the intrinsic state 
for the $J^\pm_n$ state.

\subsection{intrinsic structure}
\begin{figure}
\noindent
\epsfxsize=0.5\textwidth
\centerline{\epsffile{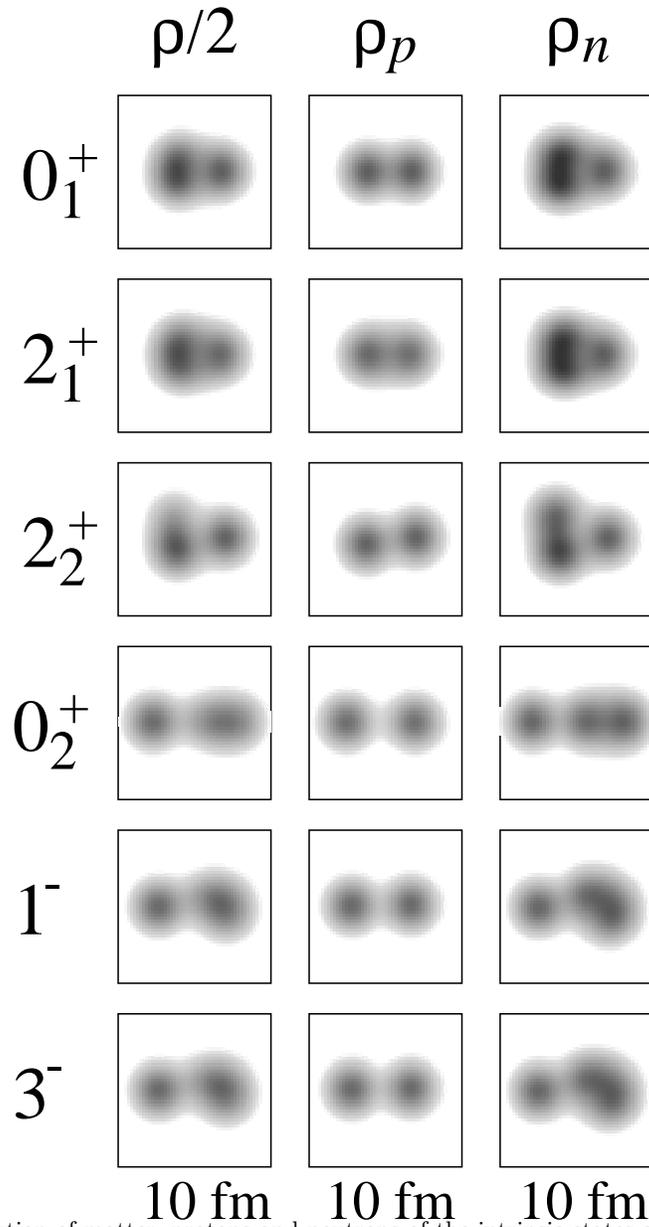}}
\caption{\label{fig:be10dens}
The density distribution of matter, protons and neutrons
of the intrinsic states are shown at left, middle and right, respectively.
The density is integrated along the axis perpendicular to adequate planes.
The figures are for the results with the interactions case(1).
}
\end{figure}

In the excited states, various kinds of structures are found.
Here we analyze the structures of the intrinsic states
$\Phi_{AMD}({\bf Z}^{J\pm}_{n})$.
It is found that the excited levels are classified into rotational bands
as $0^+_1$, $2^+_1$, $4^+_1$ states in $K^\pi=0^+_1$ band, 
$2^+_2$, $3^+_1$ in $K^\pi=2^+$ band, $0^+_2$, $2^+_3$, $4^+_2$, $6^+$
in the second $K^\pi=0^+_2$ band and $1^-$, $2^-$, $3^-$, $4^-$, $5^-$ 
in $K^\pi=1^-$ band. Particularly the 
molecule-like states with  
the well-developed 2$\alpha$ cores construct the rotational bands
$K^\pm=0^+_2$ and $1^-$ in which the level spacing is small because of the 
large moments of inertia.
The density distributions of matter, 
protons and neutrons in the intrinsic states
$\Phi_{AMD}({\bf Z}^{J\pm}_{n})$ are presented in Fig.\ \ref{fig:be10dens}.
We found the $2\alpha+2n$ structures in most of the intrinsic states.
The density of protons indicates that the clustering structure develops 
more largely in $1^-$ than $0^+_1$ and most remarkably in $0^+_2$. 
As seen in Fig.\ \ref{fig:be10dens} the intrinsic structure of $0^+_2$ 
state has an axial symmetric 
linear shape with the largest deformation, while 
$1^-$ state has an axial asymmetric shape because 
of the valence neutrons.
The structures of $0^+_1$ and $1^-_1$ states are similar to 
the ones of the previous work with the simplest version of 
AMD \cite{ENYOdoc}. The increase of the degree of  
the deformation along $0^+_1$, $1^-$ and $0^+_2$ is consistent
with the previous works such as \cite{DOTE,OGAWA,OERTZEN}. 
In the $K^\pm=1^-$ band, the deformation toward the prolate shape 
shrinks as the total spin $J$ increases.

In the $K^\pi=0^+_1$ band,
the $2\alpha$ cores weaken with the increase of the total spin
due to the spin-orbit force. The reduction of the clustering structure
is more rapid in the case of interaction (2) with the stronger spin-orbit
force and the $2^+_1$ and $4^+_1$ states in the case (2) interactions
contain the dissociation of $\alpha$.
Regarding the dissociation of the $\alpha$ cores,
the structures of those states are sensitive to the strength of the 
spin-orbit force.

The $2^+_3$ has the remarkably developed molecule-like structure 
with an exotic shape and belongs to the rotational band $K=0^+_2$ . 
On the other hand, 
the lowest $3^+$ state of $^{10}$B is the ordinary state in terms of 
$p$-shell.
The theory predicts a small value of 
the Gamow-Teller strength $^{10}$B($3^+)\rightarrow ^{10}$Be($2^+_3$)
because of the difference between the structures of the initial and final 
states.


\subsection{behavior of valence neutrons}

Even though we did not assume the existence of any clusters in the
model, we have found the $2\alpha+2n$ structures 
in most of the intrinsic states as mentioned above.
We study the behavior of the valence neutrons surrounding $2\alpha$
by analyzing the single-particle wave functions
to understand the role of the valence neutrons in the neutron-rich Be nuclei.
Considering that an intrinsic state is written by a Slater determinant
$\Phi_{AMD}$, 
the single-particle wave functions and the single-particle energies 
of an intrinsic state are determined 
by diagonalizing the single-particle Hamiltonian
by the analogy with Hartree-Fock theory.
First we transform the set of single-particle wave functions $\varphi_i$ 
of an AMD wavefunction into an orthonormal base $\tilde\varphi_\alpha$.
The single-particle Hamiltonian
can be constructed by use of the orthonormal base as follows
\cite{DOTE,ENYOdoc};
\begin{eqnarray}
h_{\alpha\beta}&=&
\langle\tilde\varphi_\alpha|\hat t|\tilde\varphi_\beta\rangle
+\sum^A_{\gamma}\langle\tilde\varphi_\alpha\tilde\varphi_\gamma|\hat v|
\tilde\varphi_\beta\tilde\varphi_\gamma-
\tilde\varphi_\gamma\tilde\varphi_\beta\rangle \nonumber\\
&+&{1\over 2}\sum^A_{\gamma,\delta}
\langle \tilde\varphi_\alpha\tilde\varphi_\gamma\tilde\varphi_\delta
|\hat v_3|\tilde\varphi_\beta\tilde\varphi_\gamma\tilde\varphi_\delta
+\tilde\varphi_\delta\tilde\varphi_\beta\tilde\varphi_\gamma
+\tilde\varphi_\gamma\tilde\varphi_\delta\tilde\varphi_\beta
-\tilde\varphi_\beta\tilde\varphi_\delta\tilde\varphi_\gamma
-\tilde\varphi_\gamma\tilde\varphi_\beta\tilde\varphi_\delta
-\tilde\varphi_\delta\tilde\varphi_\gamma\tilde\varphi_\beta\rangle,
\end{eqnarray}
where the Hamiltonian operator is
 written by a sum of kinetic term, two-body interaction term 
and three-body interaction term; $H=\sum_{i}\hat t+\sum_{i<j} \hat v_2
+\sum_{i<j<k} \hat v_3$.

\begin{figure}
\noindent
\epsfxsize=1.0\textwidth
\centerline{\epsffile{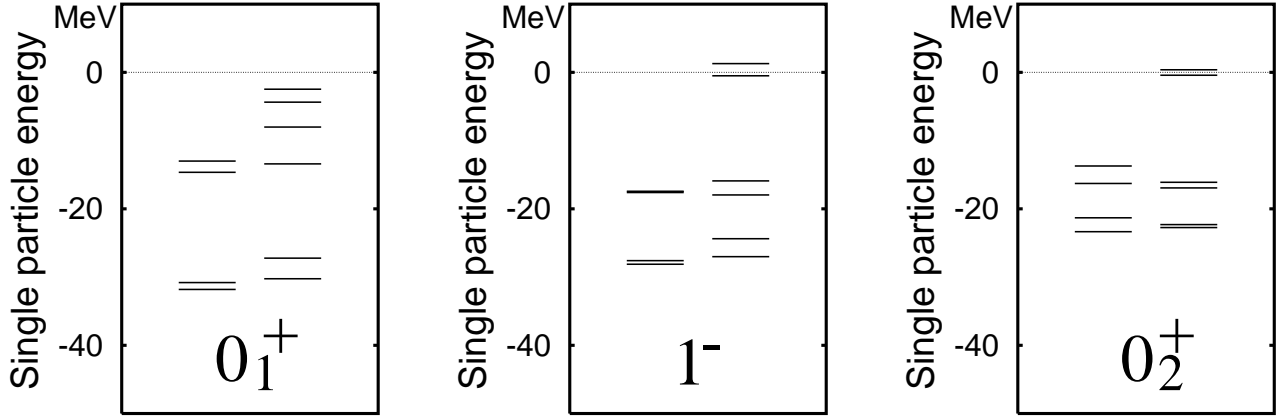}}
\caption{\label{fig:be10hfe}
Single-particle energies in the intrinsic system of the 
$0^+_1$, $1^-$ and $0^+_2$ states. The energies of protons(neutrons)
are displayed in the left(right) side in each figure.
}
\end{figure}

\begin{figure}
\noindent
\epsfxsize=0.5\textwidth
\centerline{\epsffile{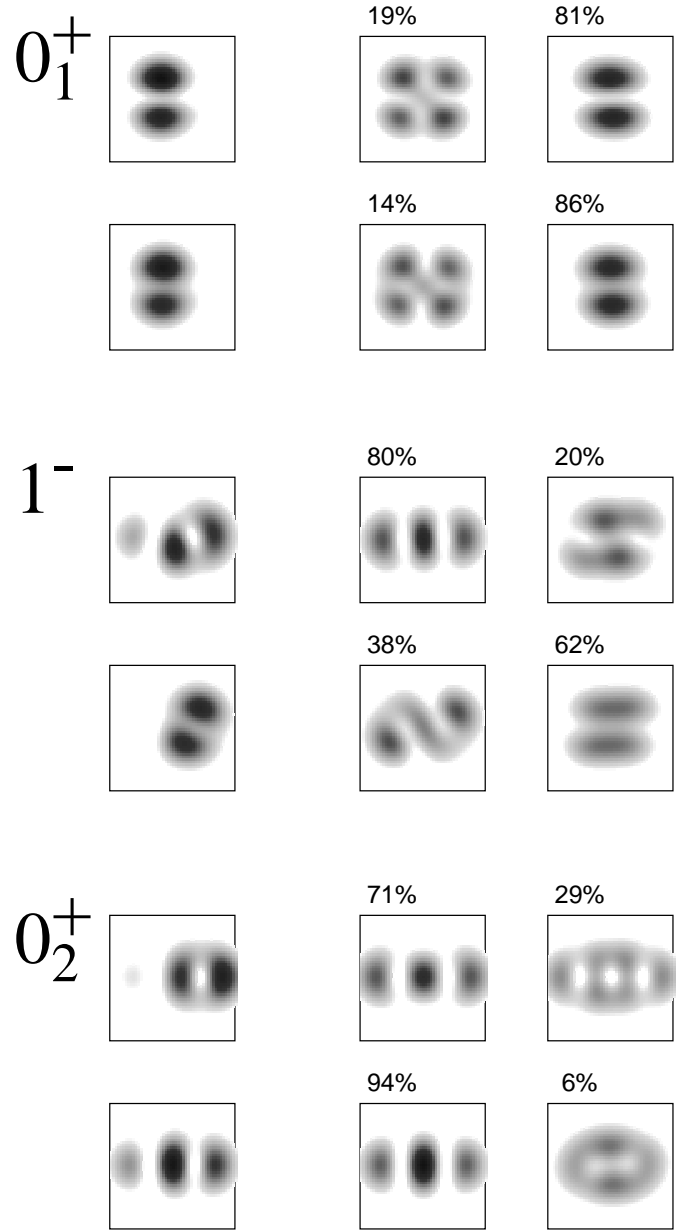}}
\caption{\label{fig:be10sing}
Density distribution of single-particle wave functions of the 
valence two neutrons
in the intrinsic states of $0^+_1$, $1^-$ and $0^+_2$
(left column).
The method to extract the single-particle wave functions are 
explained in the text.
The middle and right columns are for the density of the positive parity 
and negative parity components 
projected from the single-particle wave functions, 
respectively. The wave functions projected 
into the parity eigen states are 
normalized for presentation.
}
\end{figure}

\begin{figure}
\noindent
\epsfxsize=0.5\textwidth
\centerline{\epsffile{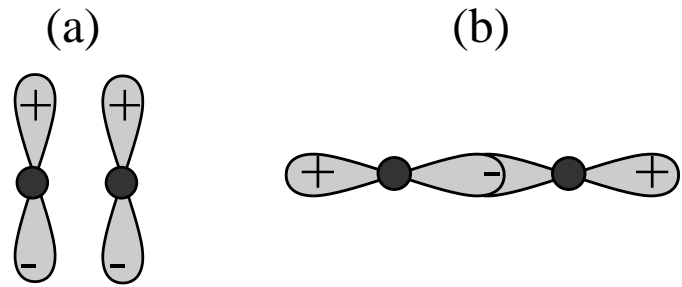}}
\caption{\label{fig:sigmapi}
Schematic figures of the molecular orbits $\pi$ bond (a)
and $\sigma$ bond (b) surrounding
2$\alpha$ clusters.
}
\end{figure}

The single-particle energies in the 
$0^+_1$, $1^-$ and $0^+_2$ states are shown 
in Fig.\ \ref{fig:be10hfe}. In each state the neutrons in the four 
orbits from the bottom correspond to the neutrons in the 
$2\alpha$ clusters.
The level spacing of these four lower orbits becomes 
smaller in $1^-$ than in $0^+_1$ and smallest in the 
$0^+_2$ state with the increase of the distance between clusters.
We consider the last two neutrons in the higher orbits as valence neutrons
surrounding 2$\alpha$ cores.
We display the density distributions of 
single-particle wave functions for the two valence neutrons
in the left column of Fig.\ \ref{fig:be10sing}. 
Figures in the middle and right columns 
of Fig.\ \ref{fig:be10sing} 
are for the normalized density of the positive 
and the negative parity eigen states 
projected from the single-particle wave functions, 
respectively.
By analyzing the single-particle wave functions it is found that
two valence neutrons of the $0^+_1$ states contain the 
negative parity components more than $80\%$
(the right column in figure\ \ref{fig:be10sing}) which seem to be 
$\pi$ bonds(Fig.\ \ref{fig:sigmapi}(a))
in terms of molecular orbits.
On the other hand in the case of $0^+_2$ band, the last 
2 neutrons are predominantly in the positive parity orbits, 
which are analogous 
to the $\sigma$ bonds (Fig.\ \ref{fig:sigmapi} (b)).
In the $1^-$ band, each valence neutron contains both the positive parity 
component like $\sigma$ and the negative parity one similar to $\pi$.
Since the parity of the total system is negative 
in the $1^-$ band,
the states after the parity projection have
one neutron in $\sigma$ orbit and the other neutron in $\pi$ orbit.
Roughly speaking, the $0^+_1$, $1^-$ and $0^+_2$ states
are understood as $2\alpha$ and two valence neutrons in $\pi^2$,
$\sigma\pi$ and $\sigma^2$ orbits, respectively. 
The interesting point is that the valence neutrons play important roles to 
develop the clustering structure in the excited bands.
We can argue that the clustering develops in the $1^-$ band 
and mostly in $0^+_2$ band owing to the $\sigma$ orbits of 
the valence neutrons, because the $\sigma$ orbit prefers the prolately 
deformed system as to gain its kinetic energy.
This idea originates from the application of 
the two centers shell model to the dimer model for Be isotopes 
by W. von Oertzen \cite{OERTZEN}
and consistent with the argument in the 
work with the method of AMD+HF by Dot\'e et al. \cite{DOTE}. 


\subsection{mechanism of clustering development} 

As mentioned above, one of the viewpoints for
the mechanism of the clustering development is the 
molecule-like orbits $\sigma$ and $\pi$.
Here we try to understand the mechanism from the other viewpoint
of the two-center clustering model.

The reason for the clustering development 
in the ordinary nuclei has been
understood as the system gains the kinetic energy 
with the development of clustering
although the potential energy looses. 
In order to understand the mechanism of clustering 
in light unstable nuclei, we think it is helpful 
to investigate the competition of the 
kinetic and the potential energies in the molecule-like states
of $^{10}$Be.

\begin{figure}
\noindent
\epsfxsize=0.7\textwidth
\centerline{\epsffile{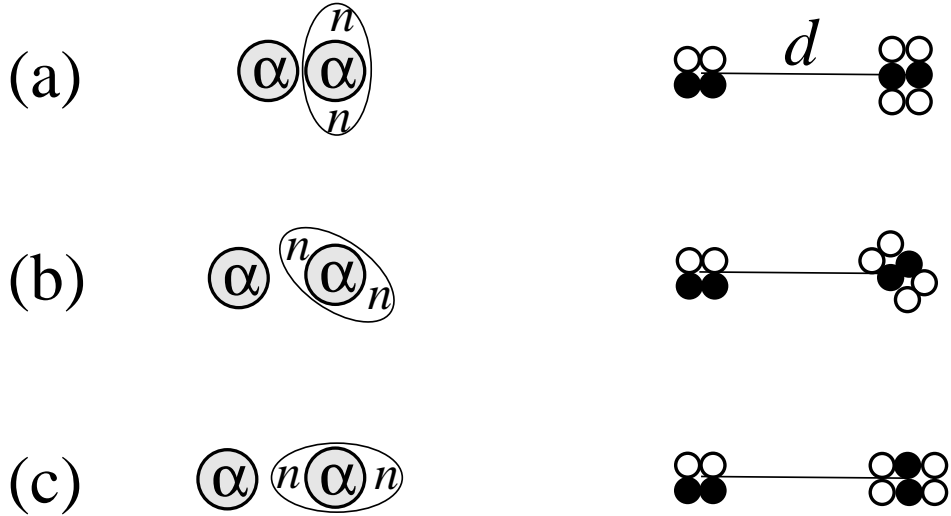}}
\caption{\label{fig:be10dim}
Schematic figures for the intrinsic structure of $0^+_1$, $1^-$ and
$0^+_2$ are shown in the left columns of (a), (b) and (c), respectively.
Right columns indicate the three kinds of 
configurations for the centers of Gaussians in the
simplified AMD wavefunctions 
which correspond to the two-center $^6$He+$\alpha$ cluster model
for the excited states $0^+_1$(a), $1^-$(b) and
$0^+_2$(c) of $^{10}$Be.
The black(white) circles correspond to the centers of Gaussians
of the single-particle wave functions for protons(neutrons).
}
\end{figure}

\begin{figure}
\noindent
\epsfxsize=.35\textwidth
\centerline{\epsffile{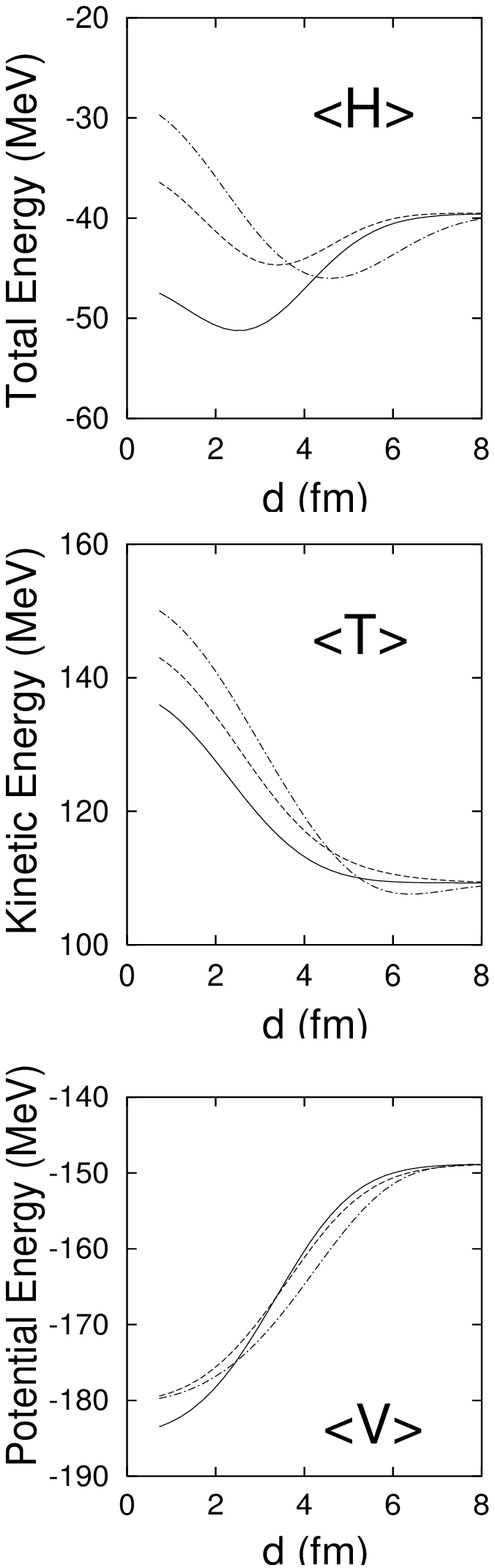}}
\caption{\label{fig:be10htv}
Total, kinetic and potential energies as the distance between clusters
in the simplified $^6$He+$\alpha$ cluster models for the excited states of 
$^{10}$Be.
The adopted interaction is the MV1 force with $m=0.62$, and 
the spin-orbit and the Coulomb forces are omitted.
The distance $d$ between clusters and the configuration for the 
$0^+_1$, $1^-$ and $0^+_2$ states are defined in the text.
The true lines, the dashed lines and the dot-dashed lines 
correspond to the energies
in the system for $0^+_1$, $1^-$ and $0^+_2$ states of $^{10}$Be, 
respectively. 
}
\end{figure}

In the density distribution of VAP results, the  
structures of the $0^+_1$, $1^-$ and $0^+_2$ states seem to be
the two-center clustering structures which consist of $^6$He+$\alpha$
(see figure \ref{fig:be10dim}). 
To estimate the dependence of the kinetic and potential energies on the
degree of the spatial clustering development we represent the three kinds of 
configurations for the $^6$He+$\alpha$ system
 corresponding to the $0^+_1$, $1^-$ and $0^+_2$ states of $^{10}$Be 
by the simplified AMD wave functions $\Phi_{AMD}({\bf Z})$ 
as follows. The intrinsic spin of the single-particle wave 
functions are fixed to be up or down
for simplicity.
For the $^6$He+$\alpha$ system with the inter-cluster distance $d$ (fm), 
the centers of single-particle Gaussian wave functions are located around  
two points $\vec{a}_1=(-3d/5\sqrt{\nu},0,0)$ and 
$\vec{a}_2=(2d/5\sqrt{\nu},0,0)$.
The three kinds of configurations of the centers for 
$0^+_1$, $1^-$ and $0^+_2$ states are 
shown in the right figures of Fig.\ \ref{fig:be10dim}. 
We put the centers for $p\uparrow$, $p\downarrow$, $n\uparrow$, 
$n\downarrow$ at the point $\vec{a}_1$ and $p\uparrow$, $p\downarrow$ at 
$\vec{a}_2$.
The centers for the last 4 neutrons are located at the points very close
to $\vec{a}_2$ as $\vec{a}_2\pm\vec{\delta}$ where enough small 
$\vec{\delta}$ is chosen so that the angle $\theta$ between $\vec{\delta}$
and $\vec{a}_2$ is $\theta$=
$\pi/2$, $\pi/4$ and $0$ corresponding to the $0^+_1$, $1^-$ 
and $0^+_2$ states, respectively. 
For the parity eigen states projected from 
these three kinds of the simplified AMD wave functions 
$\Phi_{AMD}({\bf Z})$
we calculate the expectation values of total, kinetic and potential 
energies as the function of the inter-cluster distance $d$;
\begin{eqnarray} 
& &
\langle H \rangle\equiv 
{\langle(1\pm P)\Phi_{AMD}({\bf Z})|H|(1\pm P)\Phi_{AMD}({\bf Z})\rangle
\over \langle(1\pm P)\Phi_{AMD}({\bf Z})|(1\pm P)\Phi_{AMD}({\bf Z})\rangle}\\
& &
\langle T \rangle\equiv 
{\langle(1\pm P)\Phi_{AMD}({\bf Z})|T|(1\pm P)\Phi_{AMD}({\bf Z})\rangle
\over \langle(1\pm P)\Phi_{AMD}({\bf Z})|(1\pm P)\Phi_{AMD}({\bf Z})\rangle}\\
& &
\langle V \rangle\equiv 
{\langle(1\pm P)\Phi_{AMD}({\bf Z})|V|(1\pm P)\Phi_{AMD}({\bf Z})\rangle
\over \langle(1\pm P)\Phi_{AMD}({\bf Z})|(1\pm P)\Phi_{AMD}({\bf Z})\rangle},
\end{eqnarray}
where we omit the spin-orbit and the Coulomb forces for simplicity. 
In Fig.\ \ref{fig:be10htv} we present 
the total energy, the kinetic energy and the potential energy
for the three kinds of clustering states 
$0^+_1$, $1^-$ and $0^+_2$ as the function of the distance $d$ between
$^6$He and $\alpha$ clusters. As shown in the figure for 
the total energy $\langle H \rangle$,
 the optimum distances indicate that
the clustering structure develops in the system for the $1^-$ state
and is most remarkable in the state for $0^+_2$, which is consistent with the
present results of VAP calculations.
The shift of the minimum point of the total energy is
understood by the energetically advantage of the kinetic part as follows.   
It is found that the kinetic energies in the small $d$ region
are sensitive to the configurations,
while in the case of the potential energies significant differences 
are not seen in the three configurations.
When the cluster approaches each other 
the kinetic energy in the configuration for the $1^-$ state
becomes larger than $0^+_1$ by $1/2\hbar\omega$ (about 7 MeV for $\nu=0.17$),
because all the neutrons in $0^+_1$ 
are in $0s$ and $0p$ shells, while 
in the negative parity state for $1^-$ one valence neutron  
must rise to the higher $sd$ shell in the small $d$ region.   
Since in the small $d$ limit the wave function is almost same as 
the harmonic oscillator shell model wave function due to the 
antisymmetrization, the kinetic energy of 
the $sd$ shell is larger than by $1/2\hbar\omega$ than the one of $p$ shell. 
Thus the system for $1^-$ loses as to the kinetic 
energy in the small distance $d$, 
as a result, the minimum point of total energy shift to the larger 
$d$ region than the one in the $0^+_1$ state.
In the case of the linear configuration for the excited $0^+_2$ state, 
the two valence neutrons occupy the $sd$ orbit in the small $d$ limit and
the kinetic energy is larger than $0^+_1$ by $\hbar\omega$. That is why
the optimum point $d$ in $0^+_2$ is the largest of the three.
In other words, when the cluster approaches each other, the system feels
the repulsive force in kinetic part because of the Pauli principle. 
That is the reason why the clustering structure remarkably develops in the 
$0^+_2$ state. 
In the analysis with this simplified two cluster model, we can conclude
that the clustering develops so as to gain the kinetic energy.
It is compatible with the viewpoint of the molecular $\sigma$ orbit.

\section{Summary}
\label{sec:summary}
We studied the structures of excited states of $^{10}$Be
by performing variational calculations after the spin-parity projection
in the framework of antisymmetrized molecular dynamics (AMD).
We explained the formulation of the variation after projection 
and the method to construct the higher excited states with AMD.

The excitation energies of many levels of $^{10}$Be 
were reproduced by AMD calculations
with the finite range interactions.
The theoretical results well agree to the experimental data of
$E1$, $E2$ transition strength.
The results successfully describe the new data of $B(GT)$ for 
the $\beta$ transition strength reduced from the
$0^\circ$ angle cross sections of the charge exchange reactions
$^{10}$B(t,$^3$He)$^{10}$Be$^*$. 

We discuss the structures of excited states. 
By analyzing the intrinsic structures it is found that the excited levels 
are classified into the rotational bands 
as $K^\pi=0^+_1$, $2^+$, $0^+_2$ and $1^-$.
Although the $\alpha$ clusters are not assumed in the model, 
the theoretical results show that the 2$\alpha+2n$ structures 
appear in most of the states. 
Particularly the molecular-like states with the 
developed 2$\alpha$ structure constructs the rotational bands
$K^\pi=0^+_2$ and $1^-$ in which the level spacing is small 
because of the large deformations.
The molecule-like structure develops 
more largely in $K^\pi=1^-$ than in $K^\pi=0^+_1$ and most 
remarkably in $K^\pi=0^+_2$.

We extracted the single-particle wave functions to discuss the behavior
of the valence neutrons. In the analysis of the intrinsic states of 
$0^+_1$, $1^-$ and $0^+_2$ states (the band heads of the 
rotational bands $K^\pi=0^+_1,1^-,0^+_2$), we found that 
the two valence neutrons are in the molecular orbits surrounding $2\alpha$.
The positive and negative parts of the orbits are 
analogous to the $\sigma$ and $\pi$ bonds in terms of the electron orbits in
molecules. Roughly speaking, 
the $0^+_1$ states, $1^-$ and $0^+_2$ bands 
is regarded as the states with the two valence neutrons in $\pi^2$,
$\sigma\pi$ and $\sigma^2$ orbits, respectively. 
The neutrons in the $\sigma$ orbit plays as important role for 
the development of clustering because $\sigma$ orbit profits from the 
kinetic energy in the prolately deformed system.

In order to understand the mechanism of the clustering development 
in the excited states of $^{10}$Be, we analyze the kinetic and 
the potential parts of the total energies 
with the simplified model of the two-center 
$^6$He+$\alpha$ cluster model. In the $1^-$ and $0^+_2$ states,
when the distance between clusters becomes small 
the systems feel the repulsive kinetic energy because the valence neutrons
must rise into higher orbits.
Because of the repulsive kinetic energy 
in the small inter-cluster distance
the clustering structure develops in the excited states 
$1^-$ and $0^+_2$ to gain as to the kinetic energy.

\acknowledgments
The authors would like to thank Dr. N. Itagaki for many discussions.
They are also thankful to Professor W. Von Oertzen for helpful discussions 
and comments. Valuable comments of Professor M. Fujiwara are also acknowledged.
The computational calculations of this work are supported by 
Research Center for Nuclear Physics in Osaka University,
Yukawa Institute for Theoretical Physics in Kyoto University and
the Institute of Physical and Chemical Research.


\section*{References}
\begin{thebibliography}{9}
  
\bibitem{SEYA}
 M. Seya, M. Kohno, and S. Nagata, Prog. Theor. Phys.
 {\bf 65}, 204 (1981).
\bibitem{ENYOb}
 Y. Kanada-En'yo, A. Ono, and H. Horiuchi,
Phys. Rev. C {\bf 52}, 628 (1995).
\bibitem{ENYOc}
 Y. Kanada-En'yo and H. Horiuchi,
Phys. Rev. C {\bf 52}, 647 (1995).
\bibitem{TAKAMI}
S. Takami, K. Yabana, and K. Ikeda, Prog. Theor. Phys. {\bf 96}
407 (1996); {\bf 94}, 1011 (1995). 
\bibitem{DOTE}
A. Dot\'{e}, H. Horiuchi, and Y. Kanada-En'yo, 
Phys. Rev. C {\bf 56}, 1844 (1997).
\bibitem{ITAGAKI}
N. Itagaki and S. Okabe, preprint RIKEN-AF-NP-314 (1999).
\bibitem{OGAWA}
Y. Ogawa et al., private communications.
\bibitem{OERTZEN}
W. von Oertzen, Z. Phys. A {\bf 354}, 37 (1996).
\bibitem{FUJIWARA}
I. Daito, et al., Phys. Lett. B {\bf 418} 27 (1998).
\bibitem{ONOa}
 A. Ono, H. Horiuchi, T. Maruyama, and A. Ohnishi, Prog.
 Theor. Phys. {\bf 87}, 1185 (1992).
\bibitem{ENYOa}
 Y. Kanada-En'yo and H. Horiuchi, Prog. Theor. Phys.
 {\bf 93}, 115 (1995).
\bibitem{ENYOdoc}
Y. Kanada-En'yo, Docter thesis (1996).
\bibitem{ENYOe}
 Y. Kanada-En'yo,
Phys. Rev. Lett. {\bf 81}, 5291 (1998).
\bibitem{ENYOf}
 Y. Kanada-En'yo, H. Horiuchi and A. Dot\'{e},
J. Phys. G, Nucl. Part. Phys. {\bf 24} 1499 (1998).
\bibitem{AJZEN}
F. Ajzenberg-Selove, Nucl. Phys. A {\bf 490}, 1 (1988).
\bibitem{NTAB}
{\it Table of Isotopes}, edited by C. M. Lederer and V. S. Shirley 
(Wiley, New York, 1978)
\bibitem{TOHSAKI}
 T. Ando, K.Ikeda, and A. Tohsaki, Prog. Theor. Phys.
 {\bf 64}, 1608 (1980).
\bibitem{LS}
 N. Yamaguchi, T. Kasahara, S. Nagata, and Y. Akaishi,
 Prog. Theor. Phys. {\bf 62}, 1018 (1979);
 R. Tamagaki, Prog. Theor. Phys. {\bf 39}, 91 (1968).
\end{thebibliography}
\end{document}